\documentclass[11pt]{article}
\usepackage{amsfonts}
\usepackage{amsmath,amssymb}
\usepackage{pstricks}
\usepackage{pst-node}
\usepackage{pst-tree}
\usepackage{pst-all}

\baselineskip \advance \textheight by \topskip \textwidth 470pt
\makeatletter \@addtoreset{equation}{section}

\makeatother
\def\Z{\mathbb Z}

\begin{document}

\def\arraystretch{1.2}

\title{Hidden nonlinear $su(2|2)$  superunitary symmetry of $N=2$ superextended 1D Dirac delta potential problem}
\author{\textsf{Francisco Correa$^1$}, \textsf{Luis-Miguel Nieto$^2$},
\textsf{Mikhail S. Plyushchay$^1$}\\
%EndAName
{\small \textit{$^{1}$Departamento de F\'{\i}sica, Universidad de
Santiago de Chile, Casilla 307, Santiago 2, Chile}}\\
 {\small
\textit{$^2$Departamento de F\'{\i}sica Te\'orica, At\'omica y
\'Optica,
Universidad de Valladolid, 47071, Valladolid, Spain}}\\
 {\small
\textit{E-mails: fco.correa.s@gmail.com,
luismi@metodos.fam.cie.uva.es, mplyushc@lauca.usach.cl}} }
\date{}
\maketitle

\begin{abstract}
We show that the $N=2$ superextended 1D quantum Dirac delta
potential problem is characterized by the hidden nonlinear
$su(2|2)$ superunitary symmetry. The unexpected feature of this
simple supersymmetric system is that it admits three different
$\mathbb Z_2$-gradings, which produce a separation of 16 integrals
of motion into three different sets of 8 bosonic and 8 fermionic
operators. These three different graded sets of integrals generate
two different nonlinear, deformed forms of $su(2|2)$, in which the
Hamiltonian plays a role of a multiplicative central charge. On
the ground state, the nonlinear superalgebra is reduced to the two
distinct 2D Euclidean analogs of a superextended Poincar\'e
algebra used earlier in the literature for investigation of
spontaneous supersymmetry breaking. We indicate that the observed
exotic supersymmetric structure with three different $\mathbb
Z_2$-gradings can be useful for the search of hidden symmetries in
some other quantum systems, in particular, related to the Lam\'e
equation.

\end{abstract}

\section{\protect\bigskip Introduction}

The Dirac delta potential plays a prominent role in modeling diverse
physical systems and phenomena \cite{Yang,Hag,Man,Jack}. Recently,
it was observed that the simplest \emph{bosonic }1D quantum Dirac
potential problem is characterized by a hidden \emph{nonlocal} $N=2$
supersymmetry in which a \emph{reflection} plays a role of the
grading operator \cite{CoPl}. The corresponding supersymmetry is
exact in the case of the attractive potential and is spontaneously
broken in the repulsive case.

A priori it is clear that if the system is superextended by
introduction of the discrete (spin) degrees of freedom, its explicit
supersymmetry should  enlarge. The purpose of the present paper is
to investigate how the explicit $N=2$ supersymmetry of the system
unifying the attractive and repulsive potential cases is enlarged by
the nonlocal hidden supersymmetry.

We show that the $N=2$ superextended 1D quantum Dirac potential
problem is characterized  by a nonlinear $su(2|2)$  superunitary
symmetry. The unexpected feature of this simple supersymmetric
system is that it admits three different possible choices of the
grading operator $\Gamma$, a usual one given by $\Gamma=\sigma_3$,
and two additional given by $\Gamma=R$ and $\Gamma=R\sigma_3$,
where $R$ is a reflection operator. These three choices provide us
with three different separations of the total set of 16 integrals
of motion into 8 bosonic and 8 fermionic operators. The
nonlinearity of the superalgebra formed by the integrals of motion
is similar to the nonlinearity of the symmetry algebra appearing
in the Kepler problem and associated there with a hidden symmetry
provided by the Laplace-Runge-Lenz vector \cite{Boer}. We find
that the three choices for the grading operator  give rise to the
two different forms of nonlinear, deformed $su(2|2)$
supersymmetry, with the Hamiltonian playing the role of the
multiplicative central charge of the superalgebra. It is worth to
note here that for $\Gamma=R\sigma_3$, the usual supercharges
(\ref{SU}) play a role of the \emph{bosonic} generators of the
superalgebra, whose form in this case is the same as for the
choice $\Gamma=\sigma_3$ which identifies (\ref{SU}) as
\emph{fermionic} integrals.

 On the ground state, the two types of
superalgebras are reduced to the $2D$ Euclidean versions of the
two essentially different cases of a nonstandard  superextension
of the Poincar\'e algebra, which was used earlier by Gershun and
Tkach in investigation of spontaneous supersymmetry breaking
\cite{GT}.

The paper is organized as follows. In Section 2 we identify the
set of integrals of motion of the system, and the form of the
nonlinear superalgebra generated by them depending on the choice
of the grading operator. In Section 3 we consider the action of
the integrals of motion on the energy eigenstates. In Section 4 we
identify the nature of the nonlinear supersymmetry of the system.
In Section 5  we investigate its reduction on the ground state.
Finally, section 6 is devoted to the discussion and concluding
remarks.

\bigskip

\section{Integrals of motion and grading operators}

Consider an $N=2$ supersymmetric extension of the $1D$ quantum Dirac
potential problem described by the Hamiltonian
\begin{equation}
H=p^{2}+\beta ^{2}+2\beta \delta (x)\sigma _{3}.  \label{Hamil}
\end{equation}
We choose the units $\hbar=c=1$, and put the mass of the particle
$m=\frac{1}{2}$. In this units, coordinate, momentum, energy and the
real parameter $\beta>0$ are dimensionless.
 The system is constituted by the two supersymmetric
partners corresponding to the attractive (lower component) and the
repulsive (upper component) potentials. A usual, the explicit
supersymmetry of the system (\ref{Hamil}),
\begin{equation}
\left\{ Q_{a},Q_{b}\right\} =2H\delta _{ab},\qquad \lbrack
Q_{a},H]=0~,\quad a=1,2,  \label{sup2}
\end{equation}%
is generated by the supercharges
\begin{equation}
Q_{1}= p\sigma _{1}+\beta \varepsilon (x)\sigma _{2},\qquad
Q_{2}=i\sigma _{3}Q_{1}. \label{SU}
\end{equation}
Here $\varepsilon (x)$ is the sign function,
 $\varepsilon (x)=+ 1$
($-1$) for $x>0$ ($x<0$) and $\varepsilon (0)=0$,
$\frac{d}{dx}\varepsilon (x)=2\delta (x)$, which can be treated,
e.g.,  as a limit $\varepsilon (x)=\lim_{\lambda\rightarrow
\infty}\tanh\lambda x$, and then
$\delta(x)=\lim_{\lambda\rightarrow\infty}\lambda/\cosh^2\lambda x$.
The identities for the Pauli matrices, $\sigma _{j}\sigma
_{k}=\delta _{jk}+i\epsilon _{jkl}\sigma _{l}$,  give an equivalent
representation for (\ref{SU}),
\begin{equation}
Q_{1}=\sigma _{1}\left( p+i\beta \varepsilon (x)\sigma
_{3}\right),\qquad Q_{2}=-\sigma _{2}\left( p+i\beta \varepsilon
(x)\sigma _{3}\right) . \label{SU1}
\end{equation}%

For the $N=2$ superalgebra (\ref{sup2}), a $\Z_2$-grading operator
$\Gamma$,
\begin{equation}\label{Grad1}
    \Gamma^2=1,\qquad [\Gamma,H]=0,\qquad \{\Gamma,Q_a\}=0,
\end{equation}
is identified usually with the diagonal $\sigma$-matrix,
\begin{equation}\label{G1}
    \Gamma=\sigma _{3}.
\end{equation}
 This choice  is not a unique, however, and the identification of the
 grading operator with the reflection,
\begin{equation}
     \Gamma=R,\label{G2}
\end{equation}
$R\psi(x)=\psi(-x)$, is also consistent here with relations
(\ref{Grad1}).

In accordance with \cite{CoPl}, the system has also the integrals of
motion
\begin{equation}
\tilde{Q}_{1}=p+i\beta \varepsilon (x)R\sigma _{3},\qquad
\tilde{Q}_{2}=iR\tilde{ Q}_{1},  \label{SH}
\end{equation}
which, like $R$, are nonlocal operators. Taking into account the
identity $\delta(x)R=\delta(x)$, one finds that like usual
supercharges (\ref{SU}), they are square root of the Hamiltonian,
$\tilde{Q}_{1}^2=\tilde{Q}_{2}^2=H$.  Integrals (\ref{SH}) are the
$\Z_2$-odd operators (supercharges)   for the choice (\ref{G2}), but
have to be treated as $\Z_2$-even operators for (\ref{G1}). In the
first case, they generate another copy of the $N=2$ superalgebra.

There is a third possibility for identification of the grading
operator,
\begin{equation}\label{G3}
    \Gamma=R\sigma_3.
\end{equation}
With respect to this grading, operators $\tilde{Q}_a$ are odd, while
$Q_a$ are even.

Let us see now what is the whole set of even and odd integrals of
motion of the system  for each of the three different grading
operators we have just specified.

\subsection{Grading $\boldsymbol{\Gamma=R}$}

Since with respect to (\ref{G2}) both sets of integrals (\ref{SU})
and (\ref{SH}) are odd operators, let us fix this grading and
analyze further the symmetries of the system. With respect to the
other two gradings, one of the two sets of integrals (\ref{SU}) and
(\ref{SH}) should be treated as odd, and the other as even. These
alternative choices of the grading  will be considered separately,
but we shall see finally that in all the three cases the structure
of the resulting complete supersymmetry is the same, modulo the
Hamiltonian.

For the grading (\ref{G2}), the anticommutators between the
integrals (\ref{SU}) and (\ref{SH}) should be calculated. This
gives
\begin{equation}\label{QtQ}
    \{Q_a,\tilde{Q}_1\}=2S_a,\qquad
    \{Q_a,\tilde{Q}_2\}=0,\qquad a=1,2,
\end{equation}
where
\begin{equation}\label{Sa}
    S_1=\sigma_1H-\beta\varepsilon(x)Q_2(1+R),\qquad
    S_2=i\sigma_3S_1.
\end{equation}
 Hermitian operators  (\ref{Sa}) should be
treated as new even integrals of motion. Then, the process has to be
continued: it is necessary to calculate the commutators of these
even integrals of motion between themselves and with the odd
integrals, etc. As a result, we obtain the complete list of odd
($F_1,\ldots,\, F_8 $) and even ($H$, $R$, $\Sigma_1$, $\Sigma_2$,
$B_1,\ldots\, B_4$) integrals of motion, which can be represented in
terms of the basic `building block  operators' $Q_1$, $\tilde{Q}_1$,
$S_1$, $\sigma_3$ and $R$. They are shown in Table 1.

\begin{table}[h]
\begin{center}
\begin{tabular}{|c|c|c|c|c|}
\hline
\textbf{Fermionic } & $F_{1}=Q_{1}$ & $F_{2}=R\sigma _{3}Q_{1}$ & $%
F_{3}=Q_{2}=i\sigma _{3}Q_{1}$ & $F_{4}=iRQ_{1}$ \\ \cline{2-5}
\textbf{integrals} & $F_{5}=iR\tilde{Q}_{1}=\tilde{Q}_{2}$ & $F_{6}=iR\sigma _{3}%
\tilde{Q}_{1}$ & $F_{7}=\sigma _{3}\tilde{Q}_{1}$ &
$F_{8}=\tilde{Q}_{1}$
\\ \hline\hline
\textbf{Bosonic}  & $H$ & $\Gamma =R$ & $\Sigma _{1}=\sigma _{3}$ & $%
\Sigma _{2}=R\sigma _{3}$ \\ \cline{2-5}
\multicolumn{1}{|c|}{\textbf{integrals}} & $B_{1}=Q_{1}\tilde{Q}_{1}=S_{1}$ & $%
B_{2}=i\sigma _{3}S_{1}=S_{2}$ & $B_{3}=RS_{1}$ & $B_{4}=iR\sigma
_{3}S_{1}$ \\ \hline
\end{tabular}
\end{center}
\caption{Integrals of motion, grading $\Gamma=R$}\label{T1}
\end{table}

The anticommutation relations between the fermionic integrals are
presented in Table 2.

\begin{table}[h]
\centering%
\begin{tabular}{|c||c|c|c|c|c|c|c|c|}
\hline & $F_{1}$ & $F_{2}$ & $F_{3}$ & $F_{4}$ & $F_{5}$ &
$F_{6}$ & $F_{7}$ & $F_{8}$ \\
\hline\hline
$F_{1}$ & $2H$ & $2\Sigma _{2}H$ & $0$ & $0$ & $0$ & $2B_{4}H^{\lambda }$ & $%
0$ & $2B_{1}H^{\lambda }$ \\ \hline $F_{2}$ & $2\Sigma _{2}H$ & $2H$
& $0$ & $0$ & $-2B_{2}H^{\lambda }$ & $0$ & $-2B_{3}H^{\lambda }$ &
$0$ \\ \hline $F_{3}$ & $0$ & $0$ & $2H$ & $2\Sigma
_{2}H$ & $0$ & $-2B_{3}H^{\lambda }$ & $0$ & $2B_{2}H^{\lambda }$ \\
\hline
$F_{4}$ & $0$ & $0$ & $2\Sigma _{2}H$ & $2H$ & $2B_{1}H^{\lambda }$ & $0$ & $%
-2B_{4}H^{\lambda }$ & $0$ \\ \hline
$F_{5}$ & $0$ & $-2B_{2}H^{\lambda }$ & $0$ & $2B_{1}H^{\lambda }$ & $%
2H^{1+\lambda }$ & $2\Sigma _{1}H^{1+\lambda }$ & $0$ & $0$ \\
\hline
$F_{6}$ & $2B_{4}H^{\lambda }$ & $0$ & $-2B_{3}H^{\lambda }$ & $0$ & $%
2\Sigma _{1}H^{1+\lambda }$ & $2H^{1+\lambda }$ & $0$ & $0$ \\
\hline
$F_{7}$ & $0$ & $-2B_{3}H^{\lambda }$ & $0$ & $-2B_{4}H^{\lambda }$ & $0$ & $%
0$ & $2H^{1+\lambda }$ & $2\Sigma _{1}H^{1+\lambda }$ \\ \hline
$F_{8}$ & $2B_{1}H^{\lambda }$ & $0$ & $2B_{2}H^{\lambda }$ & $0$ &
$0$ & $0$ & $2\Sigma _{1}H^{1+\lambda }$ & $2H^{1+\lambda }$ \\
\hline
\end{tabular}
\caption{Fermion-fermion anticommutation relations}\label{T2}
\end{table}

The commutation relations between bosonic integrals and between
bosonic and fermionic operators are listed in Tables 3 and 4.

\begin{table}[h]
\centering%
\begin{tabular}{|c||c|c|c|c|c|c|c|c|}
\hline & $F_{1}$ & $F_{2}$ & $F_{3}$ & $F_{4}$ & $F_{5}$ &
$F_{6}$ & $F_{7}$ & $F_{8}$ \\
\hline\hline
$\Gamma$ & $-2iF_{4}$ & $-2iF_{3}$ & $2iF_{2}$ & $2iF_{1}$ & $2iF_{8}$ & $%
2iF_{7}$ & $-2iF_{6}$ & $-2iF_{5}$ \\ \hline $\Sigma _{1}$ &
$-2iF_{3}$ & $-2iF_{4}$ & $2iF_{1}$ & $2iF_{2}$ & $0$ & $0$ & $0$ &
$0$ \\ \hline $\Sigma _{2}$ & $0$ & $0$ & $0$ & $0$ & $2iF_{7}$ &
$2iF_{8}$ & $-2iF_{5}$ & $-2iF_{6}$ \\ \hline $B_{1}$ & $0$ &
$2iF_{6}H^{1-\lambda }$ & $-2iF_{7}H^{1-\lambda }$ & $0$ & $0 $ &
$-2iF_{2}H$ & $2iF_{3}H$ & $0$ \\ \hline $B_{2}$ &
$2iF_{7}H^{1-\lambda }$ & $0$ & $0$ &
$2iF_{6}H^{1-\lambda }$ & $0$ & $-2iF_{4}H$ & $-2iF_{1}H$ & $0$ \\
\hline
$B_{3}$ & $-2iF_{5}H^{1-\lambda }$ & $0$ & $0$ & $2iF_{8}H^{1-\lambda }$ & $%
2iF_{1}H$ & $0$ & $0$ & $-2iF_{4}H$ \\ \hline
$B_{4}$ & $0$ & $-2iF_{8}H^{1-\lambda }$ & $-2iF_{5}H^{1-\lambda }$ & $0$ & $%
2iF_{3}H$ & $0$ & $0$ & $2iF_{2}H$ \\ \hline
\end{tabular}
\caption{Boson-fermion commutation relations}\label{T3}
\end{table}

\begin{table}[h]
\centering%
\begin{tabular}{|c||c|c|c|c|c|c|}
\hline & $\Sigma _{1}$ & $\Sigma _{2}$ & $B_{1}$ & $B_{2}$ &
$B_{3}$ & $B_{4}$ \\
\hline\hline $\Sigma _{1}$ & $0$ & $0$ & $-2iB_{2}$ & $2iB_{1}$ &
$-2iB_{4}$ & $2iB_{3}$
\\ \hline
$\Sigma _{2}$ & $0$ & $0$ & $-2iB_{4}$ & $2iB_{3}$ & $-2iB_{2}$ &
$2iB_{1}$
\\ \hline
$B_{1}$ & $2iB_{2}$ & $2iB_{4}$ & $0$ & $-2i\Sigma _{1}H^{2-\lambda
}$ & $0$ & $-2i\Sigma _{2}H^{2-\lambda }$ \\ \hline
$B_{2}$ & $-2iB_{1}$ & $-2iB_{3}$ & $2i\Sigma _{1}H^{2-\lambda }$ & $0$ & $%
2i\Sigma _{2}H^{2-\lambda }$ & $0$ \\ \hline $B_{3}$ & $2iB_{4}$ &
$2iB_{2}$ & $0$ & $-2i\Sigma _{2}H^{2-\lambda }$ & $0$ & $-2i\Sigma
_{1}H^{2-\lambda }$ \\ \hline
$B_{4}$ & $-2iB_{3}$ & $-2iB_{1}$ & $2i\Sigma _{2}H^{2-\lambda }$ & $0$ & $%
2i\Sigma _{1}H^{2-\lambda }$ & $0$ \\ \hline
\end{tabular}
\caption{Boson-boson commutation relations}\label{T4}
\end{table}

Here we have introduced the parameter $\lambda$ to include also
the other two cases of $\Z_2$-grading. In the present case of
$\Gamma=R$, $\lambda=0$. Due to the special choice of the indices
in the definition of fermionic operators $F$ and bosonic operators
$B$, the tables of (anti)commutation relations have a simple block
form. The Hamiltonian $H$ commutes with all the fermionic and
bosonic operators. The grading operator $\Gamma=R$ commutes with
all the bosonic integrals.

\subsection{Grading $\boldsymbol{\Gamma=\sigma_3}$}

For the choice of the $\mathbb Z_2$-grading operator
$\Gamma=\sigma_3$, as we noted, the integrals (\ref{SU}) are,
again, the fermionic operators, but the integrals (\ref{SH}) are
even operators. As a result, now the operators $S_a$ are odd
operators appearing from the commutation relations of (\ref{SU})
with (\ref{SH}). The identification of the fermionic and bosonic
integrals of motion for this case is presented in Table 5.

\begin{table}[h!]
\begin{center}
\begin{tabular}{|c|c|c|c|c|}
\hline \textbf{Fermionic } & $F_{1}=Q_{1}$ & $F_{2}=-R\sigma
_{3}Q_{1}$ & $F_{3}=-iRQ_{1}$ & $F_{4}=Q_{2}=i\sigma _{3}Q_{1}$ \\
\cline{2-5}
\multicolumn{1}{|c|}{\textbf{integrals}} & $F_{5}=RS_{1}$ & $F_{6}=-S_{1}$ & $%
F_{7}=i\sigma _{3}S_{1}$ & $F_{8}=-iR\sigma _{3}S_{1}$ \\
\hline\hline
\textbf{Bosonic}  & $H$ & $\Sigma _{1}=-R$ & $\Gamma =\sigma _{3}$ & $%
\Sigma _{2}=-R\sigma _{3}$ \\ \cline{2-5}
\multicolumn{1}{|c|}{\textbf{integrals}} & $B_{1}=-iR\sigma _{3}\tilde{Q}_{1}$ & $%
~B_{2}=-\sigma _{3}\tilde{Q}_{1}$ & $~B_{3}=-iR\tilde{Q}_{1}$ & $B_{4}=-%
\tilde{Q}_{1}$ \\ \hline
\end{tabular}
\end{center}
\caption{Integrals of motion,  grading $\Gamma
=\protect\sigma_{3}$}\label{T5}
\end{table}

These operators satisfy the (anti)commutation relations of the
same form as in the previous case, but now with  $\lambda=1$.

\subsection{Grading $\boldsymbol{\Gamma=R\sigma_3}$}

The choice $\Gamma=R\sigma_3$ is similar to the choice (\ref{G1}).
The difference with the previous case is that now the integrals
(\ref{SU}) are identified as even operators, while the integrals
(\ref{SH}) have a nature of odd operators. The integrals  $S_a$,
together with two other integrals related to them,  play again here
the role of odd operators. The identification of the fermionic and
bosonic operators is given in Table 6.

\begin{table}[h!]
\begin{center}
\begin{tabular}{|c|c|c|c|c|}
\hline
\textbf{Fermionic } & $F_{1}=\tilde{Q}_{1}$ & $F_{2}=-\sigma _{3}\tilde{Q}_{1}$ & $%
F_{3}=-iR\tilde{Q}_{1}$ & $~F_{4}=iR\sigma _{3}\tilde{Q}_{1}$ \\
\cline{2-5}
\multicolumn{1}{|c|}{\textbf{integrals}} & $F_{5}=RS_{1}$ & $~F_{6}=-S_{1}$ & $%
F_{7}=iR\sigma _{3}S_{1}$ & $F_{8}=-i\sigma _{3}S_{1}$ \\
\hline\hline
\textbf{Bosonic}  & $H$ & $\Sigma _{1}=-R$ & $\Sigma _{2}=-\sigma _{3}$ & $%
\Gamma =R\sigma _{3}$ \\ \cline{2-5}
\multicolumn{1}{|c|}{\textbf{integrals}} & $B_{1}=-i\sigma _{3}Q_{1}$ & $%
~~B_{2}=-R\sigma _{3}Q_{1}$ & $~B_{3}=-iRQ_{1}$ & $B_{4}=-Q_{1}$ \\
\hline
\end{tabular}
\end{center}
\caption{Integrals of motion, grading $\Gamma =R\protect\sigma
_{3}$}\label{T6}
\end{table}

These operators satisfy the (anti)commutation relations exactly of
the same form as for grading (\ref{G1}), i.e. the parameter
$\lambda$ is assigned here the value $\lambda=1$.

\subsection{Identification of operators}

The identification of the fermionic and bosonic operators has been
realized by us in a special way which guarantees the same form
(modulo $H$) of the (anti)commutation relations between them for the
three possible choices of the $\Z_2$-grading operator. The
identification we use corresponds to the following simple procedure.

\begin{description}
\item[Step 1.] Choose the grading operator $\Gamma$  from the set of the three Hermitian
operators
\begin{equation}\label{GGG}
    \{R,\quad \sigma_3,\quad R\sigma_3\}.
\end{equation}

\item[Step 2.] Select any  Hermitian fermionic operator $F_{1}$,
$\{F_{1},\Gamma \}=0$, from the set of the eight odd integrals
with the following properties: $\{ F_{1},F_{1}\} =2H$ and $[\Sigma
_{2},F_{1}]=0$. Here we denote by $\Sigma _{2}\neq \Gamma$ a
bosonic operator from the set (\ref{GGG}) which commutes with
$F_{1}$. The third operator from (\ref{GGG}) is denoted by $\Sigma
_{1}$. The operators $\Sigma _{1}$ and $\Sigma_2$ are defined up
to a sign.

\item[Step 3.] With the commutation relations from Table 3  of $F_{1}$
with $\Gamma$ and  $\Sigma _{1}$, the fermionic operators $F_{2}$,
$F_{3}$ and $F_{4}$ are fixed.

\item[Step 4.] Choose any other fermionic operator $F_{5}$ that satisfies
$\left\{ F_{1},F_{5}\right\} =0$ and $\left\{ F_{5},F_{5}\right\}
=2H^{1+\lambda}$.

\item[Step 5.] Repeat Step 3, but changing $F_{1}\rightarrow F_{5}$
and $\Sigma _{1}\rightarrow \Sigma _{2}$ to obtain $F_{6},~F_{7}$
and $F_{8}$.

\item[Step 6.] With the anticonmutators of the Tables 2  we identify the bosonic
operators $B_{1}$, $B_{2}$, $B_{3}$ and $B_{4}$.
\end{description}

\section{\protect\bigskip The action of the integrals of motion on
the energy eigenstates}

We have three types of the `basic' integrals of motion, $Q_{a}$,
$\tilde{Q}_{a}$, $S_{a}$, in terms of which other integrals are
obtained by multiplication with the even integrals $R$ and
$\sigma_3$. Here we will trace out the  difference between the
nature of these operators by applying them to the physical states
which we choose finally to be the eigenfunctions  of the operators
$H$, $R$ and $\sigma_3$.

In the \emph{repulsive} case (the upper component subsystem), the
Hamiltonian is $H=p^{2}+2\beta \delta (x)+\beta ^{2}$,  and the
eigenfunctions with energy $E=k^2+\beta^2>\beta ^{2}$, $k>0$, are
given by
\begin{equation}
\psi _{k}^{+}(x)=(e^{ikx}+re^{-ikx})\Theta (-x)+te^{ikx}\Theta
(x),\qquad \psi _{k}^{-}(x)=\psi _{k}^{+}(-x),
\end{equation}
where $\Theta(x)$ is the Heaviside step function.
 These wavefunctions correspond to scattering states with
the plane waves incoming, respectively, from $-\infty $ and
$+\infty$, with reflection and transmission coefficients given by
\cite{CoPl}
\begin{equation}
r=\frac{\beta }{ik-\beta },\qquad t=\frac{ik}{ik-\beta },\qquad
k=\sqrt{E-\beta ^{2}}.
\end{equation}
Let us construct from these states the eigenfunctions of  a fixed
parity,
\begin{equation}
\psi^{(+)}_{k,+}(x)=\frac{(ik-\beta )}{2i}(\psi _{k}^{+}+\psi
_{k}^{-})=k\cos kx+\beta \varepsilon (x)\sin kx,
\end{equation}%
\begin{equation}
\psi^{(+)}_{k,-}(x)=\frac{1}{2i}(\psi _{k}^{+}-\psi _{k}^{-})=\sin
kx,
\end{equation}
$R\psi^{(+)}_{k,\pm}(x)=\pm\psi^{(+)}_{k,\pm}(x)$.

The \emph{attractive} (lower component) case is obtained by the
change $\beta \rightarrow -\beta $ in all the previous formulae,
$H=p^{2}-2\beta \delta (x)+\beta ^{2}$,
\begin{equation}
\psi ^{(-)}_{k,+}(x)=k\cos kx-\beta \varepsilon (x)\sin kx, \qquad
\psi^{(-)}_{k,-}(x)=\sin kx.
\end{equation}
In addition to the scattering states, this subsystem has a unique
bound state described by a normalized wave function,
\begin{equation}
\psi^{(-)}_{0}(x)=\sqrt{\beta }e^{-\beta |x|}.  \label{psid0}
\end{equation}
Due to a special value of the constant term in the Hamiltonian, the
energy of this bound state is equal to zero.

Summarizing, for the system (\ref{Hamil}) we have the scattering
eigenstates,
\begin{equation}
\Psi _{k,+}^{(+)}(x)=\left(
\begin{array}{c}
k\cos kx+\beta \varepsilon (x)\sin kx \\
0
\end{array}
\right) ,\quad \Psi _{k,-}^{(+)}(x)=\sqrt{k^{2}+\beta ^{2}}\left(
\begin{array}{c}
\sin kx \\
0
\end{array}
\right),
\end{equation}
\begin{equation}
\Psi _{k,+}^{(-)}(x)=\left(
\begin{array}{c}
0 \\
k\cos kx-\beta \varepsilon (x)\sin kx%
\end{array}
\right) ,\quad \Psi _{k,-}^{(-)}(x)=\sqrt{k^{2}+\beta ^{2}}\left(
\begin{array}{c}
0 \\
\sin kx
\end{array}
\right),
\end{equation}
and the bound state of zero energy,
\begin{equation}
\Psi _{0}^{(-)}(x)=\left(
\begin{array}{c}
0 \\
\sqrt{\beta }e^{-\beta |x|}\
\end{array}
\right).
\end{equation}
We have introduced here a convenient normalization to  simplify the
relations that follow.

The ordinary  supercharges (\ref{SU}) change the \emph{parity} and
the \emph{upper-lower} subspaces of the eigenfunctions,
\begin{equation}
Q_{1}\Psi _{\pm }^{(\pm )}=\pm i\sqrt{k^{2}+\beta ^{2}}\Psi _{\mp
}^{(\mp )}, \qquad
 Q_{1}\Psi _{\mp }^{(\pm )}=\mp i\sqrt{k^{2}+\beta
^{2}}\Psi _{\pm }^{(\mp )},
\end{equation}
\begin{equation}
Q_{2}\Psi _{\pm }^{(\pm )}=\sqrt{k^{2}+\beta ^{2}}\Psi _{\mp }^{(\mp
)}, \qquad Q_{2}\Psi _{\mp }^{(\pm )}=-\sqrt{k^{2}+\beta ^{2}}\Psi
_{\pm }^{(\mp )},
\end{equation}
where for simplicity we have omitted the index $k$.
 The supercharges of
the \emph{hidden} supersymmetry (\ref{SH}) change only the
\textit{parity},
\begin{equation}
\tilde{Q}_{1}\Psi _{\pm }^{(\alpha )}=\pm i\sqrt{k^{2}+\beta
^{2}}\Psi _{\mp }^{(\alpha )},\qquad \tilde{Q}_{2}\Psi _{\pm
}^{(\alpha )}=\sqrt{k^{2}+\beta ^{2}}\Psi _{\mp }^{(\alpha )},\qquad
\alpha =+,-,
\end{equation}
The $S_{a}$ operators change only the \textit{upper-lower subspaces}
of eigenfunctions,
\begin{equation}
S_{1}\Psi _{\alpha }^{(\pm )}=(k^{2}+\beta ^{2})\Psi _{\alpha
}^{(\mp )},\qquad S_{2}\Psi _{\alpha }^{(\pm )}=\pm i(k^{2}+\beta
^{2})\Psi _{\alpha }^{(\mp )},\qquad \alpha =+,-.
\end{equation}
The unique bound state is annihilated by the integrals of motion
$Q_a$, $\tilde{Q}_a$ and $S_a$,
\begin{equation}
Q_{a}\Psi _{0}^{(-)}=0,\qquad \tilde{Q}_{a}\Psi _{0}^{(-)}=0,\qquad
S_{a}\Psi _{0}^{(-)}=0.
\end{equation}
As a consequence, it is annihilated by all the odd integrals of
motion $F_1,\ldots,F_8$ and all the even integrals $B_1,\ldots,B_4$
(as well as by $H$).

\section{\protect\bigskip Identification of supersymmetry, generic
case}

To clarify the nature of the supersymmetry generated by the complete
set of eight fermionic and eight bosonic integrals of motion, we
define the following linear combinations of the even operators,
\begin{equation}\label{K12}
 P_{1}^{\left( \pm \right)
}=\frac{1}{4} \left( B_{1}\pm B_{3}\right)=\frac{1}{2}B_1\Pi_\pm
,\quad
 P_{2}^{\left( \pm \right) }=-\frac{1}{4}\left( B_{2}\pm
B_{4}\right)=- \frac{1}{2}B_2\Pi_\pm ,
\end{equation}
\begin{equation}\label{K3}
J_{3}^{\left( \pm \right) }=\frac{1}{4}\left( \Sigma _{1}\pm ~\Sigma
_{2}\right) =\frac{1}{2}\Sigma _{1}\Pi_\pm ,
\end{equation}
where  $\Pi_\pm=\frac{1}{2}(1\pm \Gamma)$ are the projectors. These
operators satisfy the relations
\begin{equation}\label{K1K2}
\left[ P_{1}^{\left( \pm \right) },P_{2}^{\left( \pm \right)
}\right] =iJ_{3}^{\left( \pm \right) }H^{2-\lambda },
\end{equation}
\begin{equation}\label{KaK3}
\left[ J_{3}^{\left( \pm \right) },P_{a}^{\left( \pm \right)
}\right] =i\epsilon _{ab}P_{b}^{\left( \pm \right) },\qquad a,b=1,2,
\end{equation}
\begin{equation}\label{KKi}
    [P^{(+)}_a,P^{(-)}_b]=[J_3^{(+)},P^{(-)}_a]=[J_3^{(-)},P^{(+)}_a]=
    [J_3^{(+)},J_3^{(-)}]=0.
\end{equation}
The commutation relations (\ref{K1K2}), (\ref{KaK3}) correspond to a
direct sum of two deformed $su(2)$ algebras. In particular,
relations (\ref{K1K2}) are reminiscent of the commutation relations
satisfied by the components of the Laplace-Runge-Lenz vector in the
quantum Kepler problem. Being reduced to the energy eigensubspace of
nonzero eigenvalue $E>\beta^2>0$, the rescaled operators
$J^{(\pm)}_a=P^{(\pm)}_a/E^{1-\frac{\lambda}{2}}$ and $J^{(\pm)}_3$
generate the Lie algebra $su(2)\oplus su(2)$ and satisfy the
relations $J^{(+)}_i J^{(+)}_i=\frac{3}{4}\Pi_+$, $J^{(-)}_i
J^{(-)}_i=\frac{3}{4}\Pi_-$, where the summation in $i=1,2,3$ is
assumed. These relations mean that the two common eigenstates of the
Hamiltonian with energy $E>\beta^2>0$ and of the grading operator
$\Gamma$ with eigenvalue $+1$ carry spin-1/2 representation for
$J^{(+)}_i$, and are the spin-0 states for $J^{(-)}_i$. The states
with $\Gamma=-1$ carry spin one-half representation for $J^{(-)}_i$,
and are of spin zero for $J^{(+)}_i$. Fermionic integrals mutually
transform the states from these eigenspaces of the grading operator.
In accordance with the total number of independent fermionic
generators, the energy subspace with $E>\beta^2>0$ carries an
irreducible representation of the $su(2|2)$ superunitary symmetry,
which is a  supersymmetric extension of the bosonic symmetry
$u(1)\oplus su(2)\oplus su(2)$, where the $u(1)$ subalgebra is
generated by the grading operator, see Ref. \cite{Sen}. Having in
mind that the Hamiltonian appears in a generic form of the
superalgebra as a multiplicative central charge, we conclude that
the system possesses the nonlinear $su(2|2)$ superunitary symmetry
in the sense of Refs. \cite{Boer,AIS,Para}.

In order to understand better the identified symmetry  at the
algebraic level, let us  define the following linear complex
combinations of the even operators (\ref{K12}),
$P^{(\pm)}_\pm=P^{(\pm)}_1\pm iP^{(\pm)}_2$, and of the odd
operators,
\begin{equation}\label{Fnew}
{\cal S}_1=\frac{1}{2}\left( F_{1}+ iF_{3}\right),\quad {\cal
S}_2=\frac{ 1}{2}\left( F_{2}+ iF_{4}\right) ,\quad {\cal
Q}_1=\frac{1}{2}\left( F_{7}+ iF_{5}\right) , \quad {\cal
Q}_2=\frac{1}{2}\left( F_{8}+ iF_{6}\right),
\end{equation}
and denote $\bar{\cal S}_{1,2}={\cal S}_{1,2}^\dagger$, $\bar{\cal
Q}_{1,2}={\cal Q}_{1,2}^\dagger$. In terms of these complex
combinations, the (anti)commutation relations including fermionic
integrals are presented in Tables 7 and 8.

\begin{table}[h]
\begin{tabular}{|c||c|c|c|c|c|c|c|c|}
\hline & $\mathcal{S}_{1}$ & $\mathcal{\bar{S}}_{1}$ &
$\mathcal{S}_{2}$ & $\mathcal{\bar{S}}_{2}$ & $\mathcal{Q}_{1}$ &
$\mathcal{\bar{Q}}_{1}$ & $
\mathcal{Q}_{2}$ & $\mathcal{\bar{Q}}_{2}$ \\
\hline\hline
$\mathcal{S}_{1}$ & $0$ & $H$ & $0$ & $\Sigma _{2}H$ & $0$ & $0$ & $%
2P_{-}^{(+)}H^{\lambda }$ & $2P_{-}^{(-)}H^{\lambda }$ \\ \hline
$\mathcal{\bar{S}}_{1}$ & $H$ & $0$ & $\Sigma _{2}H$ & $0$ & $0$ & $0$ & $%
2P_{+}^{(-)}H^{\lambda }$ & $2P_{+}^{(+)}H^{\lambda }$ \\ \hline
$\mathcal{S}_{2}$ & $0$ & $\Sigma _{2}H$ & $0$ & $H$ & $-2P_{-}^{(+)}H^{%
\lambda }$ & $2P_{-}^{(-)}H^{\lambda }$ & $0$ & $0$ \\ \hline
$\mathcal{\bar{S}}_{2}$ & $\Sigma _{2}H$ & $0$ & $H$ & $0$ & $%
2P_{+}^{(-)}H^{\lambda }$ & $-2P_{+}^{(+)}H^{\lambda }$ & $0$ & $0$
\\ \hline
$\mathcal{Q}_{1}$ & $0$ & $0$ & $-2P_{-}^{(+)}H^{\lambda }$ & $%
2P_{+}^{(-)}H^{\lambda }$ & $0$ & $H^{1+\lambda }$ & $0$ & $\Sigma
_{1}H^{1+\lambda }$ \\ \hline
$\mathcal{\bar{Q}}_{1}$ & $0$ & $0$ & $2P_{-}^{(-)}H^{\lambda }$ & $%
-2P_{+}^{(+)}H^{\lambda }$ & $H^{1+\lambda }$ & $0$ & $\Sigma
_{1}H^{1+\lambda }$ & $0$ \\ \hline $\mathcal{Q}_{2}$ &
$2P_{-}^{(+)}H^{\lambda }$ & $2P_{+}^{(-)}H^{\lambda }$
& $0$ & $0$ & $0$ & $\Sigma _{1}H^{1+\lambda }$ & $0$ & $H^{1+\lambda }$ \\
\hline
$\mathcal{\bar{Q}}_{2}$ & $2P_{-}^{(-)}H^{\lambda }$ & $2P_{+}^{(+)}H^{%
\lambda }$ & $0$ & $0$ & $\Sigma _{1}H^{1+\lambda }$ & $0$ &
$H^{1+\lambda }$ & $0$ \\ \hline
\end{tabular}
\caption{Fermion-fermion anticommutation relations. Here
$\Sigma_{1(2)}=2\left( J_{3}^{(+)}+(-)J_{3}^{(-)}\right)$}\label{T7}
\end{table}

\begin{table}[h]
\begin{tabular}{|c||c|c|c|c|c|c|c|c|}
\hline & ${\cal S}_{1}$ & $\mathcal{\bar{S}}_{1}$ &
$\mathcal{S}_{2}$ & $\mathcal{\bar{S}}_{2}$ & $\mathcal{Q}_{1}$ &
$\mathcal{\bar{Q}}_{1}$ & $
\mathcal{Q}_{2}$ & $\mathcal{\bar{Q}}_{2}$ \\
\hline\hline
$\Gamma $ & $-2\mathcal{S}_{2}$ & $2\mathcal{\bar{S}}_{2}$ & $-2\mathcal{S}%
_{1}$ & $2\mathcal{\bar{S}}_{1}$ & $-2\mathcal{Q}_{2}$ & $2\mathcal{\bar{Q}}%
_{2}$ & $-2\mathcal{Q}_{1}$ & $2\mathcal{\bar{Q}}_{1}$ \\ \hline
$J_{3}^{(+)}$ & $-\frac{1}{2}\mathcal{S}_{1}$ & $\frac{1}{2}\mathcal{\bar{S}}%
_{1}$ & $-\frac{1}{2}\mathcal{S}_{2}$ &
$\frac{1}{2}\mathcal{\bar{S}}_{2}$ &
$-\frac{1}{2}\mathcal{Q}_{1}$ & $\frac{1}{2}\mathcal{\bar{Q}}_{1}$ & $-\frac{%
1}{2}\mathcal{Q}_{2}$ & $\frac{1}{2}\mathcal{\bar{Q}}_{2}$ \\ \hline
$J_{3}^{(-)}$ & $-\frac{1}{2}\mathcal{S}_{1}$ & $\frac{1}{2}\mathcal{\bar{S}}%
_{1}$ & $-\frac{1}{2}\mathcal{S}_{2}$ &
$\frac{1}{2}\mathcal{\bar{S}}_{2}$ &
$\frac{1}{2}\mathcal{Q}_{1}$ & $-\frac{1}{2}\mathcal{\bar{Q}}_{1}$ & $\frac{1%
}{2}\mathcal{Q}_{2}$ & $-\frac{1}{2}\mathcal{\bar{Q}}_{2}$ \\ \hline
$P_{-}^{(+)}$ & $0$ & $-\mathcal{Q}_{1}H^{1-\lambda }$ & $0$ & $\mathcal{Q}%
_{2}H^{1-\lambda }$ & $0$ & $\mathcal{S}_{1}H$ & $0$ &
$-\mathcal{S}_{2}H$
\\ \hline
$P_{-}^{(-)}$ & $0$ & $-\mathcal{\bar{Q}}_{1}H^{1-\lambda }$ & $0$ & $-%
\mathcal{\bar{Q}}_{2}H^{1-\lambda }$ & $\mathcal{S}_{1}H$ & $0$ & $\mathcal{S%
}_{2}H$ & $0$ \\ \hline
$P_{+}^{(+)}$ & $\mathcal{\bar{Q}}_{1}H^{1-\lambda }$ & $0$ & $-\mathcal{%
\bar{Q}}_{2}H^{1-\lambda }$ & $0$ & $-\mathcal{\bar{S}}_{1}H$ & $0$ & $%
\mathcal{\bar{S}}_{2}H$ & $0$ \\ \hline
$P_{+}^{(-)}$ & $\mathcal{Q}_{1}H^{1-\lambda }$ & $0$ & $\mathcal{Q}%
_{2}H^{1-\lambda }$ & $0$ & $0$ & $-\mathcal{\bar{S}}_{1}H$ & $0$ & $-%
\mathcal{\bar{S}}_{2}H$ \\ \hline
\end{tabular}
\caption{Boson-fermion commutation relations }\label{T8}
\end{table}

{}From the (anti)commutation relations it follows that the operators
$J^{(+)}_3$, and $P^{(+)}_\pm$ act invariantly (in the sense of
commutator) on every of the spin one-half multiplets $(\bar{\cal
Q}_1,{\cal S}_1)$, $(\bar{\cal S}_1,{\cal Q}_1)$, $(\bar{\cal
Q}_2,{\cal S}_2)$, $(\bar{\cal S}_2,{\cal Q}_2)$, where the first
and second components of every multiplet are the eigenstates of
$J^{(+)}_3$ with  eigenvalues $+\frac{1}{2}$ and $-\frac{1}{2}$.
Analogous multiplets for the bosonic operators $J^{(-)}_3$, and
$P^{(-)}_\pm$ are $({\cal Q}_1,{\cal S}_1)$, $(\bar{\cal
S}_1,\bar{\cal Q}_1)$, $({\cal Q}_2,{\cal S}_2)$ and $(\bar{\cal
S}_2,\bar{\cal Q}_2)$. The eigenstates of the $u(1)$ generator
$\frac{1}{2}\Gamma$ are the linear combinations of (\ref{Fnew}) and
of the Hermitian conjugate operators: the operators $({\cal S}_1-
{\cal S}_2)$, $(\bar{\cal S}_1+ \bar{\cal S}_2)$, $({\cal Q}_1-
{\cal Q}_2)$ and $(\bar{\cal Q}_1+ \bar{\cal Q}_2)$ are the
eigenstates of the eigenvalue $+1$, while $({\cal S}_1+{\cal S}_2)$,
$(\bar{\cal S}_1- \bar{\cal S}_2)$, $({\cal Q}_1+ {\cal Q}_2)$ and
$(\bar{\cal Q}_1- \bar{\cal Q}_2)$ are the eigenstates of the
eigenvalue $-1$. The anticommutators of the odd integrals produce
linear combinations of $H$ and the  generators of the two chiral
$su(2)$ algebras modulo $H$.

\section{\protect\bigskip Supersymmetry reduced on the ground state}

As we have seen, the three possible gradings result in only two
forms of the supersymmetry algebra given by the (anti)commutation
relations with $\lambda=0$ for $\Gamma=R$, and with $\lambda=1$ for
$\Gamma=\sigma_3$ and $\Gamma=R\sigma_3$. The reduction of these
superalgebras on the subspace of scattering states with fixed energy
 $E>\beta^2>0$ effectively results, for both values of $\lambda$,  in
the same superextension of the $u(1)\oplus su(2)\oplus su(2)$ Lie
algebra. However, the essential difference between the two forms
reveals itself on the ground state of zero energy.

Indeed, with $H=0$, the bosonic part of the superalgebra is reduced
to the algebra $u(1)\oplus e(2)\oplus e(2)$, where the first term
corresponds to the integral $\Gamma$, and other two correspond to
the two copies of the 2D Euclidean algebras generated by the
rotation operator $J^{(+)}_3$, the translation generators
$P^{(+)}_a$, $[P^{(+)}_1,P^{(+)}_2]=0$, and by its analogs for the
$\Gamma=-1$ subspace.

For $\lambda=1$, in accordance  with Table \ref{T7}, all the
fermionic operators anticommute, and in this case we have also the
commutation relations of the form $[P,{\cal Q}]=0$, $[P,{\cal
S}]\sim {\cal Q}$, where we do not display the indices and do not
distinguish fermionic operators from the Hermitian conjugate.

On the other hand, for $\lambda=0$, the fermionic integrals commute
with the translation generators, $[P,{\cal S}]=[P,{\cal Q}]=0$, and
anticommute between themselves in the following way: $\{{\cal
S},{\cal S}\}=\{{\cal Q},{\cal Q}\}=0$, $\{{\cal S},{\cal Q}\}\sim
P$. Therefore, the difference between the grading $\Gamma=R$ on one
hand, and the gradings $\Gamma=\sigma_3$ and $\Gamma=R\sigma_3$ on
the other hand, reveals itself on the ground state.

It is interesting to note that the supersymmetry of the form similar
to the present supersymmetry reduced on the ground state was
analyzed by Gershun and Tkach \cite{GT} in the context of
spontaneous supersymmetry breaking in 3+1 dimensions. The
superalgebra used by them contains two parameters, $a$ and $b$, (see
Eq. (1) from \cite{GT}). The cases ($a=0$, $b\neq 0$) and ($a\neq
0$, $b=0$) correspond, respectively,  to our cases $\lambda=0$ and
$\lambda=1$.

\section{Concluding remarks and outlook}

\bigskip

Although  the system we have studied is simple, the existence of
nontrivial (nonlocal) integrals of motion reveals a new
supersymmetric structure. A part of the nontrivial integrals of
motion  of the $N=2$ superextended system comes from the hidden
supersymmetry of the corresponding bosonic (spinless) quantum
mechanical Dirac delta potential problem, while other conserved
quantities appear naturally with the addition of fermionic degrees
of freedom. Combining both supersymmetries, the usual one
generated by (\ref{SU}), and the hidden supersymmetry generated by
(\ref{SH}), we find new, a priori unexpected nontrivial (nonlocal)
integrals of motion (\ref{Sa}). As a result, we  enrich the
knowledge on the system with a new property: its supersymmetric
structure admits three different grading operators. With respect
to these three different $\mathbb Z_2$-gradings, every of the
three sets of nontrivial integrals (\ref{SU}), (\ref{SH}) and
(\ref{Sa}) (and other associated integrals obtained from them by
multiplication with grading operators) is identified once as a set
of bosonic generators, while two other sets are identified in this
case as fermionic operators.

Hidden supersymmetry exists also in a bosonic (spinless)
P\"oschl-Teller (PT) quantum problem for integer values of the
parameter of the system (for the details see ref. \cite{CoPl}),
the appropriate limit of which corresponds to the Dirac potential
problem. However, in PT problem, hidden $N=2$ supersymmetry is
nonlinear \cite{AIS,Para}, of the order defined by the value of
the parameter. Therefore, one should expect that such a hidden
nonlinear supersymmetry has also to reveal itself in the simplest
$N=2$ superextended version of the PT system. However, due to the
nonlinear character of the hidden supersymmetry, the resulting
complete supersymmetry has to have a more rich nature in
comparison with the Dirac delta potential case. Moreover, since
the PT problem at those special values of the parameter (being
reflectionless) carries some characteristics of a free particle,
and in a certain limit produces a particle model of conformal
mechanics \cite{ConfMod}, it would be interesting to investigate
the PT system from the viewpoint of possible hidden superconformal
symmetry.

Another system that reveals hidden nonlinear supersymmery is the
quantum periodic Lam\'e problem \cite{CNP}, which reproduces the
hidden supersymmetry of the PT system in a certain limit. Following
the line presented in this work,  we also are going to investigate
the question on hidden supersymmetry in the superextended Lam\'e
system, which has some remarkable properties in the context of
supersymmetry \cite{dunsuk}.

Let us note that the observed exotic supersymmetric structure with
three different $\mathbb Z_2$-gradings can be effective in the
search of hidden symmetries in some other quantum systems. In
particular, a general knowledge of this structure is helpful for
identification of the unknown hidden symmetries in the broad class
of the associated Lam\'e systems \cite{assoLame},  and in solving
them. The key point there is that to every Lam\'e system two
different algebraisation schemes are related \cite{sl2R}, with
which two different sets of the non-diagonal integrals of the type
(\ref{SU}) and (\ref{Sa}) can be naturally associated. The
detailed study of Lam\'e and P\"oschl-Teller systems in the light
of the revealed here nontrivial supersymmetric structure will be
presented elsewhere \cite{PCLpre}.

Since the reduction of the observed hidden nonlinear $su(2|2)$
supersymmetry of the $N=2$ superxtended Dirac delta potential
problem to the ground state of zero energy produces the 2D Euclidean
analogs of two particular cases of a special superextension of the
Poincar\'e symmetry \cite{GT}, it would be interesting to
investigate the origin of this latter symmetry. One could expect
that it should appear in a special (contraction) limit of some
superextension of the de Sitter symmetry.

Finally, it would be interesting to investigate the question on
existence of the hidden supersymmetry in the systems of many
particles, in particular,  with Dirac delta interactions, as well
as in nonlinear integrable systems.

\vskip 0.2cm \textbf{Acknowledgements.}  We are grateful to  Dmitri
Sorokin for valuable communications. The work has been supported
partially by the FONDECYT Projects 1050001 and 7070024, by CONICYT,
by  Spanish Ministerio de Educaci\'on y Ciencia (Project
MTM2005-09183) and Junta de Castilla y Le\'on (Excellence Project
VA013C05). FC and MP thank Department of Theoretical Physics of
Valladolid Univerisity for hospitality.

\end{document}